# Broadband temporal localization and delocalized temporal edge states in time photonic crystals

Junkai Jiang[1,2,†], Hao Hu[1,2,†,*], Yang Long[3], Liangliang Liu[1,2], Songyan Hou[2,4], Dongjue Liu[5,*] and Zhuo Li[1,2,*]

[1]National Key Laboratory of Microwave Photonics, College of Electronic and Information Engineering, Nanjing University of Aeronautics and Astronautics, Nanjing 211106, China

[2] Key Laboratory of Radar Imaging and Microwave Photonics Ministry of Education, College of Electronic and Information Engineering, Nanjing University of Aeronautics and Astronautics, Nanjing 211106, China

[3]Division of Physics and Applied Physics, School of Physical and Mathematical Sciences, Nanyang Technological University, Singapore 637371, Singapore

[4]Guangzhou Institute of Technology, Xidian University, Guangzhou 510555, China

[5]National Key Laboratory of Scattering and Radiation, Shanghai 200438, China

[†]These authors contribute equally.

[*]**Email:** hao.hu@nuaa.edu.cn (Hao Hu); dongjue001@e.ntu.edu.sg (Dongjue Liu); lizhuo@nuaa.edu.cn (Zhuo Li)

**Time photonic crystals have attracted growing attention in recent years owing to their abilities to enable broadband field enhancements, e.g., free-space electromagnetic waves, dipolar emissions, free-electron radiation, etc. While the non-Hermitian nature of time photonic crystals is primarily attributed to their dependence on external temporal modulations, the constituent materials are oftentimes assumed to be Hermitian. How the material-induced non-Hermiticity interplays with the intrinsic non-Hermitian dynamics of time photonic crystals remains rarely explored. In this work, we demonstrate that the non-Hermiticity arising from the bi-anisotropic electromagnetic response of materials introduces a new mechanism to manipulate the localization of temporal bulk and edge states in time photonic crystals. To be specific, the temporal bulk states in our configurations exhibit remarkable attenuation or amplification, which is theoretically predicted by extending the generalized Brillouin zone framework to the temporal domain. Our analysis reveals that the attenuation or amplification strength, quantified by the temporal penetration depth, is directly governed by electromagnetic constitutive parameters. By appropriately tuning these parameters, we uncover new phenomena including broadband temporal localization—the collective concentration of energy towards a certain time moment, and delocalized temporal edge states.**

In quantum mechanics, Hermitian Hamiltonians describe closed systems with conserved energy. However, most physical systems interact with their environments, leading to non-Hermitian Hamiltonians that account for external energy exchange. These non-Hermitian systems exhibit unique phenomena distinct from their Hermitian counterparts, such as non-orthogonal eigen-states [1], eigen-state coalescence at exceptional points and lines [2-4], and unconventional dynamical responses [5]. Among these, the non-Hermitian skin effect (NHSE), characterized by the exponential localization of eigen-states at the system's boundaries, has emerged as a particularly intriguing phenomenon. Under open boundary conditions, the skin effect invalidates the Bloch theorem, complicating predictions of Bloch wave behaviors, including field distributions and energy spectra.

The NHSE underpins various novel physical properties, including topological funneling [6,7], delocalized edge states [8], and entanglement suppression [9,10]. These phenomena challenge the traditional bulk-boundary correspondence, which links edge states to bulk topological invariants in Hermitian systems. To address these challenges, the generalized Brillouin zone (GBZ) and non-Bloch band theory have been developed. A new definition of GBZ (i.e. $\beta=e^{ikd}$ where $k$ is Bloch wavevector in complex plane and $d$ is the spatial period) has been proposed [11], providing frameworks to predict and qualify the localization behaviors of eigen-states. A so-called Bloch point on the intersection of the GBZ and unit circle, is further introduced to separate the amplified ($|\beta|>1$) and attenuated ($|\beta|<1$) localization of bulk eigen-states, and the bulk states are delocalized at this intersection point [12]. Furthermore, the non-Bloch bulk-boundary correspondence demonstrates that edge states in non-Hermitian systems are governed by non-Bloch topological invariants (i.e. the winding number) defined on the GBZ. Specifically, the winding number could explicitly predict the number of zero mode and the emerging position of edge states [11]. And the framework of GBZ not only applies to study topological property, but also spreads to various areas such as non-Hermitian band structures, Green's functions, parity-time (PT) symmetry, chiral damping and dynamics [5,13-20]. As a typical case, the non-Bloch exceptional points corresponding to specific values of $\beta$ on GBZ, could determine PT symmetry-breaking phase transitions [18]. Moreover, the Green's function formula defined on the contour integration pf GBZ, is proposed as an efficient formulation to design the gain and directionality of amplification [19].

Recently, time-varying media have emerged as a novel class of non-Hermitian systems, where the energy of the system is exchanged with the environment through the time modulation [21-24]. As time offers a unique degree of freedom to overcome conventional physical bounds in stationary materials, time-varying media enable a wealth of novel effects including magnetic-free nonreciprocity [24-26], localized mode extraction [27], high-efficient frequency conversion [28,29], etc. One of the most representative time-varying media is the time photonic crystal (TPC), whose constitutive parameters are periodically modulated in time [30-35]. Distinct from the energy bandgap of space photonic crystal, the bandgap of TPC is opened in momentum, so-called momentum bandgap [42]. Such a unique bandgap can amplify photons in a broad range of momenta, favorably offering broadband field enhancement for spontaneous emission [33], free electron radiation [34,35] and free-space electromagnetic waves [30,36]. Despite their non-Hermitian nature, previous work has shown that classical Bloch theorem provides sufficient accuracy to predict behaviors of eigen-states in TPCs. As a typical example, the TPC with time-inversion symmetry would demonstrate excellent bulk-edge correspondence, simply by translating the Bloch theorem from space to time domain [21]. As we will show in this work, such a translation is effective mainly because the adopted constituent materials of TPCs in a finite temporal duration are generally Hermitian even though TPCs are intrinsically non-Hermitian.

Recent advances have shown that when the constituent materials of TPCs become non-Hermitian, they can exhibit unconventional phenomena, such as topological temporal boundary states [43] and enhancing broadband absorption [44]. However, temporal analogues of non-Hermitian effects, especially the coexistence of localized bulk states and delocalized edge states, are much less explicitly addressed in the existing TPC literature. Moreover, most previous works introduce non-Hermiticity by gain-loss and asymmetric coupling [45-49]. Up to date, how non-Hermitian effects interact with bi-anisotropic response in TPCs still remains unexplored.

In this work, we demonstrate that bi-anisotropic materials can achieve TPCs with controllable non-Hermiticity. This is inspired from Ref. [8], where the author innovatively investigate wave behaviors in stationary photonic crystals made of bi-anisotropic materials. Via considering bi-anisotropic electromagnetic responses [37-41], as characterized by constitutive parameters (e.g. the chirality parameter), the constitutive material becomes non-Hermitian. The inclusion of non-Hermiticity leads to the successful prediction of non-Hermitian skin modes and delocalized edge states in spatial dimension. Our study reflects that these materials' non-Hermiticity would also lead to the temporal localization of bulk states in the TPC. This is distinct from the TPC constructed by Hermitian materials, where all the bulk states are extensively distributed in time domain [Fig. 1(a)]. To predict the temporal localization features of bulk states, we introduce the temporal version of GBZ defined in the complex plane of frequencies rather than wavevectors. This newly defined GBZ not only predicts the temporal localization direction of the bulk states (i.e., the amplification or attenuation of bulk states in the TPC), but also quantifies the corresponding localization strength (i.e., the amplification or attenuation rate). Our results further show that the amplification or attenuation rate could be flexibly controlled by the electromagnetic constitutive parameters, e.g., electromagnetic coupling coefficients. By strategic engineering of electromagnetic constitutive parameters, we demonstrate the broadband temporal localization in Fig. 1(b), where the electromagnetic field and energy within the TPC travel into a specific time moment—electromagnetic field and energy in a finite frequency band are enhanced at a specific time moment and then undergo temporal decay. The localization dynamics of time edge states are manipulated by the relative alignment of their localization vectors with those of temporally confined bulk modes. Congruent localization directions of temporal bulk and edge states enhance edge-state confinement. Conversely, opposing vector orientations suppress the localization strength of temporal edge states. Interestingly, temporal edge states initially localized at the time domain wall (red dash-line rectangle in Fig. 1) could become completely delocalized via precisely engineering the electromagnetic constitutive parameters [Fig. 1(c&d)]. The space-time dynamical description of temporal modes, especially localized bulk and delocalized edge modes, has been investigated in the appendix H.

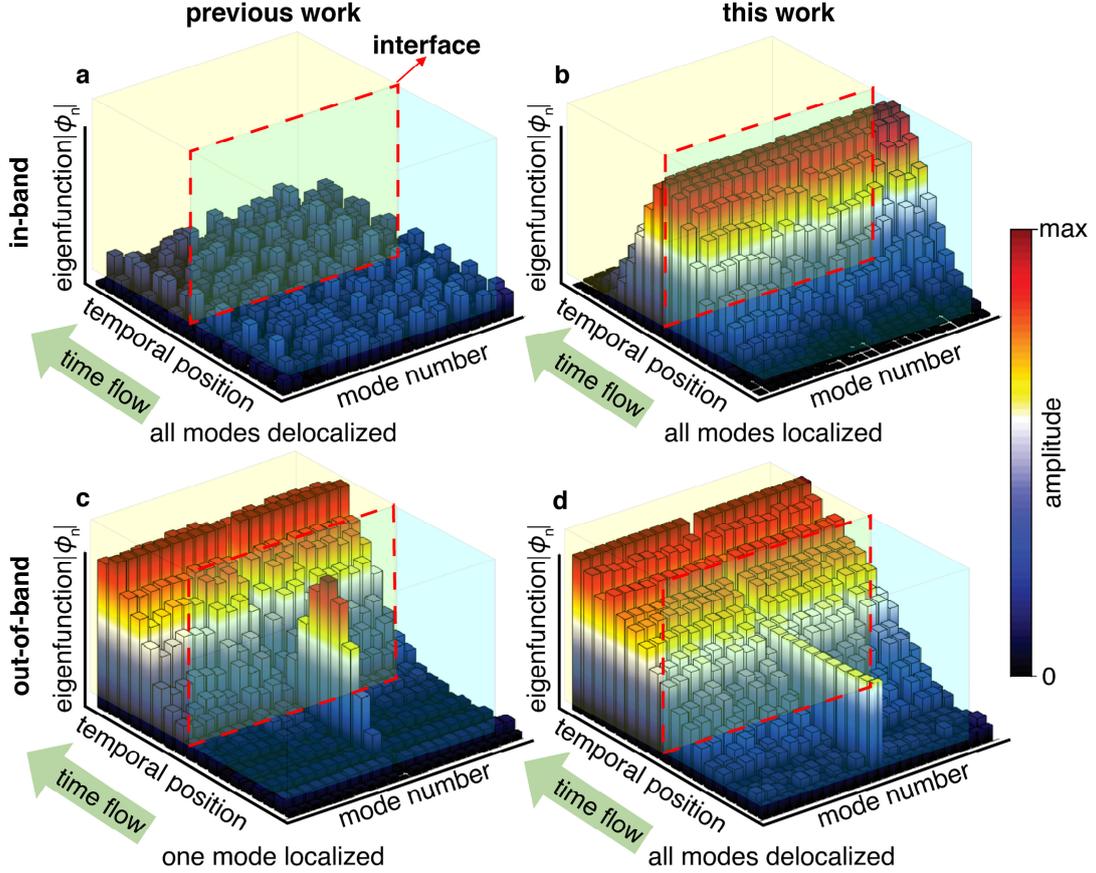

Fig. 1: Comparison of mode behaviors in TPCs made of Hermitian and non-Hermitian materials. (a) Spacetime mode in the band in a TPC purely made of Hermitian materials. (b) Spacetime mode in the band in a TPC made of non-Hermitian materials. (c) Spacetime mode out of the band in a TPC purely made of Hermitian materials. (d) Spacetime mode out of the band in a TPC made of non-Hermitian materials. The time interface between two different TPCs is remarked by the red rectangle.

Without loss of generality, we study a TPC constructed by two materials periodically alternating in time. These two materials show bi-anisotropic electromagnetic responses, i.e.,

$$D_x = \varepsilon_0 \varepsilon_{r,i} E_x + i(\kappa_i/c) H_y \tag{1a}$$

$$B_y = i(\kappa_i/c) E_x + \mu_0 H_y \tag{1b}$$

, where $c$ is the light speed in vacuum; $\varepsilon_0$ and $\mu_0$ are the permittivity and permeability in vacuum, respectively; two studied materials are denoted by subscripts $i=1,2$, respectively; $\varepsilon_{r,i}$ and $\kappa_i$ correspond to the relative permittivity and chirality parameter in the constitutive material marked by $i$, respectively; $E_x$, $D_x$, $H_y$, $B_y$ refer to the $x$-polarized electric field, $x$-polarized displacement field, $y$-polarized magnetic field, and $y$-polarized magnetic flux density, respectively. Combining Eq. (1) with Maxwell Equation, one can readily obtain the eigen-equation of the bi-anisotropic material as

$$n_i \begin{bmatrix} D_x \\ B_y \end{bmatrix} = \mathcal{M}_i \sigma_x \begin{bmatrix} D_x \\ B_y \end{bmatrix} \tag{2}$$

, where $n_i = k_i/\omega_i$ is the effective refractive index; $k$ is the wavevector; $\sigma_x$ is the first Pauli matrix; and $\mathcal{M}_i$ is a material matrix expressed as (see detailed derivation in Appendix A)

$$\mathcal{M}_i = \begin{bmatrix} c\varepsilon_0 \varepsilon_{r,i} & i\kappa_i \\ i\kappa_i & c\mu_0 \end{bmatrix}. \tag{3}$$

The material matrix shows that the constitutive materials of TPC are non-Hermitian once bi-anisotropy is considered. Specifically, without the bi-anisotropy (i.e., $\kappa = 0$), the material matrix satisfies $\mathcal{M} = \mathcal{M}^{\dagger}$. This implies that constitutive materials are Hermitian, even though the TPC as a whole is non-Hermitian. This type of non-Hermiticity intrinsically originates from the temporally modulated system. However, a different type of non-Hermiticity induced by chirality parameter (i.e., $\text{Re}(\kappa) \neq 0$) stems from breaking the constituent materials' symmetry. Specifically, when $\text{Re}(\kappa) \neq 0$, the material matrix would not satisfy $\mathcal{M} = \mathcal{M}^{\dagger}$, breaking the parity symmetry and leading to the non-Hermitian symmetries [8,46,50], i.e. the materials' non-Hermiticity. It manifests objectively at different $|\beta^{\pm}|$ (equivalently different $\text{Im}(\omega_B^{\pm})$), directly governing the emergence of complex eigen spectrum under PBC and the temporal non-Bloch GBZ behavior discussed in the manuscript. Finally, we highlight that such a unique medium satisfying the Eq. (1) can be practically realized in the platform of transmission-line metamaterials that are periodically loaded by nonreciprocal gyrators [41] (see the detailed information in appendix D).

For the TPC composed of bi-anisotropic media, the eigen-spectra and dispersion under periodic boundary condition (PBC) and open boundary condition (OBC) are markedly different. The specific dispersion equation and OBC boundary condition are given by (the derivation procedure of dispersion, OBC boundary condition and eigen-spectra are displayed in Appendix B)

$$\cos\left[(\omega_B - \omega^{\Sigma})T\right] = \frac{\eta+1}{2}\cos\left(\omega_{2(3)}^{\triangle}t_{2(3)} + \omega_1^{\triangle}t_1\right) + \frac{-\eta+1}{2}\cos\left(\omega_{2(3)}^{\triangle}t_{2(3)} - \omega_1^{\triangle}t_1\right) \quad (4)$$

and

$$|\beta^+| = |\beta^-| \text{ or } \text{Im}(\omega_B^+) = \text{Im}(\omega_B^-) \quad (5)$$

where $\omega_1^{\triangle} = (\omega_1^+ - \omega_1^-)/2$, $\omega_3^{\triangle} = (\omega_{2(3)}^+ - \omega_{2(3)}^-)/2$, and $\eta = (Z_{2(3)}/Z_1 + Z_1/Z_{2(3)})/2$.

$\omega^\Sigma T = \omega_1 t_1 + \omega_{2(3)} t_{2(3)}$, here, $\omega_1 = (\omega_1^+ + \omega_1^-)/2$, $\omega_3 = (\omega_{2(3)}^+ + \omega_{2(3)}^-)/2$; $\beta^\pm = e^{-i\omega_B^\pm T}$ is the temporal Bloch factor; $T$ is the time period of TPC; $\omega_B$ is the Bloch frequency. The superscript of plus and minus corresponds to the forward/backward propagating waves; the temporal durations of the isotropic and bi-anisotropic materials are denoted by $t_1$ and $t_{2(3)}$, respectively; $\omega_1$ and $\omega_{2(3)}$ refers to the angular frequency of the isotropic and bi-anisotropic materials, respectively; $Z_1$ and $Z_{2(3)}$ refers to the impedance of the isotropic and bi-anisotropic materials, respectively. Meanwhile, Eq. (5) determines the formation of non-Bloch temporal modes in time photonic crystals. These temporal modes, on the other hand, are the superposition of forward propagating and backward propagating waves induced by time interfaces. In other words, Eq. (5) is necessary to be satisfied such that the eigen-spectra of non-Bloch temporal modes could be obtained. As a direct application, the eigen-spectra of non-Bloch temporal modes enable us to efficiently evaluate their corresponding decays or amplification rates in time as demonstrated in Fig. 2(h) of the main text. Here, in contrast to PBC structure with an infinite number of periodic units [Fig. 2(a)], we introduce the OBC by enforcing the electromagnetic field $B_y = 0$ or $D_x = 0$ at the interface of TPCs [Fig. 2(e)]. To illustrate the differences of the eigen-spectra and dispersions in PBC and OBC, we revisit the eigen-spectra of TPC using purely Hermitian materials ($\kappa = 0$). The TPCs made of Hermitian materials under PBC and OBC demonstrate consistent eigen-spectra, where Brillouin zone is defined in the real-valued frequency (i.e., the Bloch frequency $\omega_B$) and bandgap is open up in the real-valued wavevector, i.e., momentum band-gap, as shown in Fig. 2(b&c). The real $\omega_B$ reflects that eigen-states are always extensive in time domain (i.e. delocalization in time domain) since they temporally evolve as $e^{i\omega_B t}$. Differently, in the TPCs composed of non-Hermitian materials ($\kappa \neq 0$), the eigen-spectra and dispersion relationship under PBC and OBC no longer overlap with each other. To be specific, the wavevector becomes a complex value while the Bloch frequency remains real under PBC [Fig. 2(f&g)]. This means that despite introducing non-Hermitian materials, the dispersion under PBC indicates that all the eigen-states of TPCs should be extensive over the time domain [Fig. 2(a)]. On the other hand, the wavevector is real but the Bloch frequency becomes a complex number when applying OBC [Fig. 2(f&g)]. In other words, the temporal bulk states initially extensive under PBC could become localized in the time interface under OBC [Fig. 2(e)]. The more specific discussion on how exactly the relative permittivity and chiral parameter in TPC rely on time is described in Fig. R2 of the appendix A.

To predict the temporal localization of bulk states, we extend the GBZ from the space to the time domain. To start with, we revisit the GBZ in a spatially periodic system described in the frequency domain. $\beta = e^{ikd}$ represents the complex Bloch factor, where its magnitude $|\beta|$ determines the spatial attenuation or amplification of eigenmodes, while its phase $\arg(\beta)$ corresponds to the spatial phase shift per unit cell. Inspired from the spatial counterpart, we introduce the generalized phase shift of temporal Bloch wave as $\beta = e^{-i\omega_B T}$. Here $\omega_B$ has a non-zero imaginary part corresponding to the amplification and attenuation rate in time. Then under OBC, the trajectory of the end point of $\beta$ (i.e. the coordinate of $\beta$ in the complex plane) forms a series of temporal GBZs in the complex plane, corresponding to different dispersion bands. When $\kappa = 0$ and the Bloch frequencies $\omega_B$ are real, the GBZ is always a unit circle in Fig. 2(d), implying the extensive nature of eigen-states [Fig. 2(a)]. Inversely, when the frequencies are complex, the GBZ is no longer overlapped with the unit circle. To be specific, when $\kappa < 0$ ($\kappa > 0$), the GBZ lies outside (inside) the unit circle in Fig. 2(h), corresponding to the amplification (attenuation) of eigen-states in time [Fig. 2(e)].

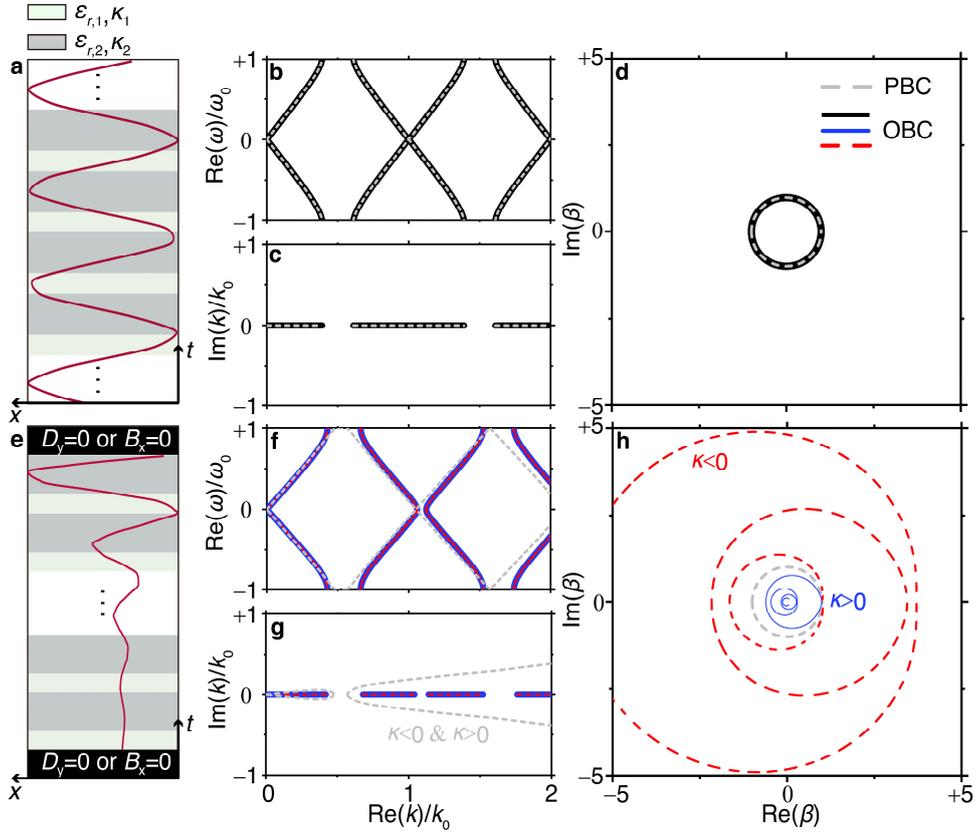

**Fig. 2: Boundary condition and band diagrams of TPCs. (a) Schematic of the TPC under PBC. (e) Schematic of the TPC under OBC. In (a&e), the red-solid curves show the distribution of temporal bulk states (illustrated by the electric displacement field $D_x$) in the TPC. (b&f) The dispersion curves under PBC and OBC for the TPC. (c&g) The eigen-spectra under PBC and OBC for the TPC.**

**(d&h) Brillouin zone and generalized Brillouin zone of the TPC.** $\beta$ **is the generalized phase shift, denoted by** $\beta=e^{i\omega_B t}$**, where** $\omega_B$ **is the Bloch frequency. In (b,c,d), the TPC is constructed by two materials with** $\kappa_1 = 0$ **and** $\kappa_2 = 0$**. In (f,g,h), the TPC is constructed by two materials with** $\kappa_1 = 0$ **and** $\kappa_2 \neq 0$**.**

The temporal localization of bulk states enables the broadband temporal localization, where the electromagnetic fields and energy automatically concentrate towards a designated time interface. To demonstrate this, we consider a time-varying structure constructed by two different TPCs, i.e., the TPC A and the TPC B. The OBC is considered in such a time-varying structure and the background material is air in Fig. 3(a). The TPC A and TPC B are made up of 10 cells. Each cell in TPC A is temporally composed of an isotropic material with relative permittivity of $\varepsilon_{r,1}$ and bi-anisotropic material with relative permittivity of $\varepsilon_{r,2}$. Each cell in TPC B is temporally composed of an isotropic material with relative permittivity of $\varepsilon_{r,1}$ and bi-anisotropic material with relative permittivity of $\varepsilon_{r,3}$. The chirality parameters of bi-anisotropic materials in TPC A and B are $\kappa_2$ and $\kappa_3$, respectively. When $\kappa_2 = -0.18$ and $\kappa_3 = 0.18$, all the bulk states in TPC A demonstrate the amplification in time while those in TPC B tend to temporally attenuate. According to the definition, the penetration depth of temporal bulk states could be qualified as (see detailed procedure derived from Appendix C)

$$\delta_{bulk} = \left| \frac{T\left(\varepsilon_{r,2(3)} + \kappa_{2(3)}^2\right)}{kc\kappa_{2(3)}t_{2(3)}} \right| \tag{6}$$

, $\varepsilon_{r,2(3)}$ and $\kappa_{2(3)}$ are respectively the relative permittivity and chirality parameter of the bi-anisotropic materials, while the other constitutive materials remain isotropic with the relative permittivity of $\varepsilon_{r,1}$. Note that in the limit of $\kappa_{2(3)} = 0$, the bulk states in TPC are always extensive over the entire time domain due to the infinite value of penetration depth, as illustrated in Fig. 3(b).

Moreover, the temporal localization of bulk states also impacts the localization of temporal edge states inside the momentum bandgap. To quantify the localization depth of temporal edge states, we derive their eigenvalues and eigenfunctions. By enforcing the time boundary at the time domain wall, we derive the dispersion relation of temporal edge states [see the derivation procedure in the Appendix B],

$$\frac{T_1^+ + R_1^+}{T_{-1}^+ + R_{-1}^+} = \frac{T_1^+ - R_1^+}{T_{-1}^+ - R_{-1}^+} \tag{7}$$

, where $T_1^+$ and $R_1^+$ correspond to the coefficients of forward and backward propagating waves in TPC B, respectively; $T_{-1}^+$ and $R_{-1}^+$ correspond to the coefficients of forward and backward propagating waves in TPC A, respectively. When $\kappa_2 = 0$ and $\kappa_3 = 0$, temporal edge states have an initial penetration depth of $\delta_0 = 1/|\mathrm{acosh}(\eta/T)|$. Introducing bi-anisotropy could either enhance or reduce the penetration depth, as indicated by (see detailed procedure derived from Appendix C)

$$\delta_{\mathrm{edge}} = \frac{1}{\dfrac{1}{\delta_0} \pm \left|\dfrac{kc\kappa_{2(3)}t_{2(3)}}{T\left(\varepsilon_{r,2(3)} + \kappa_{2(3)}^2\right)}\right|} \tag{8}$$

The choice of $\pm$ depends on the temporal localization direction of bulk modes. If the localization direction of bulk states coincides with that of edge states [Fig. 3(c)], we select the plus sign in Eq. (8) such that the temporal localization of edge states becomes even stronger than the initial one. Otherwise, the minus sign is chosen in Eq. (8), showing that the temporal localization of edge states is weakened instead [Fig. 3(d)]. Interestingly, when the chirality parameters satisfy $\kappa_2 = 0.18$ and $\kappa_3 = -0.18$, the temporal edge states at $k \cdot (cT) = 23.3356$ (i.e., $k = 2.477 k_0$, where $k_0 \cdot (cT) = 3\pi$) would demonstrate infinite penetration depth. This means that the edge states that initially localized at the time interface could become completely extensive at certain chirality parameters. To make our conclusion more general, we discuss the influence of material dispersion and absorption on behaviors of bulk and edge states in the appendix E and G, respectively. We also consider the variation of temporal duration instead of fixing each temporal slab (see details in the appendix F), finding that the behaviors of broadband temporal localization show high robustness.

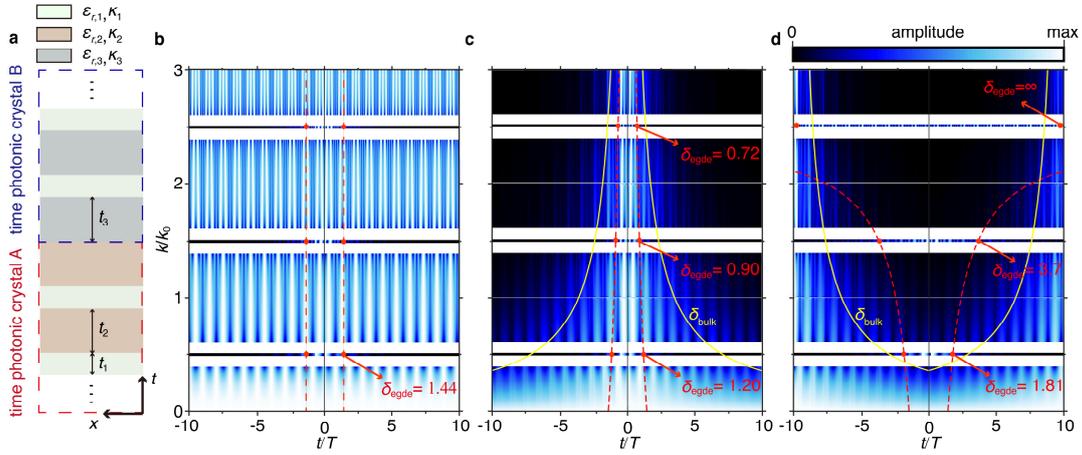

**Fig. 3: Broadband temporal localization and delocalized temporal edge states. (a) Schematic of time-varying structure made of two TPCs. (b) Field distributions of electric displacement as a function of time and wavevector in the time varying structure with $\kappa_2 = 0$ and $\kappa_3 = 0$. (c) Field distributions of electric displacement in the time varying structure with $\kappa_2 = -0.18$ and $\kappa_3 = 0.18$. (d) Field**

**distributions of electric displacement in the time varying structure with** $\kappa_2 = 0.18$ **and** $\kappa_3 = -0.18$**. In (b,c,d), the penetration depth of temporal bulk states is denoted as the yellow-solid lines, and the penetration depth of temporal edge states is denoted as the red-solid dots.**

Finally, we show that penetration depth of temporal bulk and edge states could be precisely controlled by tuning the chirality parameters in the bi-anisotropic materials. To illustrate this, we plot the eigen-function of temporal bulk and edge states in the time-varying structure at different values of chirality parameter. As $\kappa = -\kappa_2 = \kappa_3$ varies from 0 to 0.5, the decreased penetration depth of temporal bulk states makes their mode profiles in Fig. 4(a) much sharper, indicating the enhanced localization for broadband temporal localization. On the other hand, by tuning $\kappa = \kappa_2 = -\kappa_3$ from 0 to 0.5, the increased penetration depth of temporal edge states makes their mode profiles in Fig. 4(b) tend to be flat, leading to the observation of delocalized edge state. And finally the localization direction of temporal edge states is reversed. Furthermore, one can even achieve temporal edge states featuring both the extensive and localized nature. To be specific, we fix the chirality parameter $\kappa_2 = 0$ but vary $\kappa = -\kappa_3$ from 0 to 0.5. As a consequence, the temporal edge states in the TPC A remain localized, while those in the TPC B experience the delocalization transition [Fig. 4(c)]. These findings have a potential application for advanced amplifier, including dynamic switching from field attenuation and amplification or robust and tunable field enhancement at designed time moment [53]. More phenomena unique to the time boundaries have been stated on the appendix I.

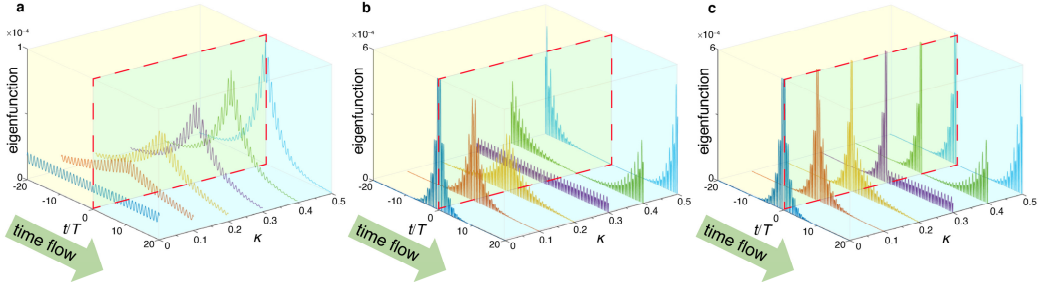

**Fig. 4: The impact of the chirality parameter $\kappa$ on the penetration depth of temporal eigen-states in time-varying structure in Fig. 3(a). (a) As $\kappa = -\kappa_2 = \kappa_3$ varies from 0 to 0.5, the localization of temporal bulk states is strengthened, leading to the enhanced broadband temporal localization. (b) As $\kappa = \kappa_2 = -\kappa_3$ varies from 0 to 0.5, the localization of temporal edge states is weakened, leading to the delocalized edge states. (c) As only $\kappa = -\kappa_3$ varies from 0 to 0.5, the localization of temporal edge states in TPC B is weakened, leading to the delocalized edge states.**

To summarize, the non-Hermiticity of constitutive materials enforces a dramatic influence on the band topology of TPCs, necessitating the extension of GBZ from spatial domain to temporal domain. This extension enables the observation of novel phenomena, including broadband temporal localization and delocalized temporal edge states in TPCs. These results provide a deeper understanding of the interplay between material properties and non-Hermitian temporal dynamics, offering new avenues for the advanced control of light-matter interactions in time photonic systems. Note that the materials' non-Hermiticity is introduced in this work only by breaking the parity symmetry in considered constitutive materials, i.e., via introducing the chirality parameter corresponding to the imaginary part of electromagnetic coupling coefficients. Alternatively, the material's non-Hermiticity can be realized by breaking the time-reversal symmetry or (both parity and time-reversal symmetry) if one also considers that the real part of electromagnetic coupling coefficients, i.e., the Tellegen or omega parameters [37,40]. Our work thus inspires future exploration of rich physics in TPC constructed by non-Hermitian materials.

**Research funding:** National Natural Science Foundation of China (Grants No. 12404363, No. 61701246); Natural Science Foundation of Jiangsu Province (Grants No. BK20241374); Qing Lan Project of Jiangsu Province (Grants No. 1004-YQR22031); Distinguished Professor Fund of Jiangsu Province; Fundamental Research Funds for the Central Universities, NUAA (Grants No. NS2024022); Shanghai Pujiang Program (Grants No. 23PJ1411500); Stable Operation Fund Project of National Key Laboratory of Scattering and Radiation (Grants No. 2023JCJQLB013050408); Natural Science Basic Research Program of Shaanxi Province (2024JC-YBQN-0682); Open Project of Key Laboratory of Radar Imaging and Microwave Photonics (Nanjing University of Aeronautics and Astronautics), Ministry of Education (NJ20240003); China Postdoctoral Science Foundation (2024M762538); Eric and Wendy Schmidt AI in Science Postdoctoral Fellowship, a Schmidt Futures program.
**Author contribution:** All authors have accepted responsibility for the entire content of this manuscript and approved its submission. H. H. conceived the original idea. J. J. performed the analytical calculations. H.H. and Z.L. supervise the project. All authors discussed the results and provided feedback. J. J., D.L. and H. H. wrote the manuscript.
**Conflict of interest:** The authors declare no conflicts of interest.
**Informed consent:** Informed consent was obtained from all individuals included in this study.
**Data availability:** Data underlying the results presented in this paper are not publicly available at this time but may be obtained from the authors upon reasonable request.

# Appendix A: The material's non-Hermiticity induced by bi-anisotropy

In this appendix, we discuss the material's non-Hermiticity in details induced by bi-anisotropy. Without loss of generality, we consider a bi-anisotropic material whose constitutive relation fulfills

$$\begin{bmatrix} D_x \\ D_y \\ D_z \end{bmatrix} = \varepsilon_0 \begin{pmatrix} \varepsilon_{r,i} & 0 & 0 \\ 0 & \varepsilon_{r,i} & 0 \\ 0 & 0 & \varepsilon_{r,i} \end{pmatrix} \begin{pmatrix} E_x \\ E_y \\ E_z \end{pmatrix} + \begin{pmatrix} 0 & \xi_i/c & 0 \\ 0 & 0 & 0 \\ 0 & 0 & 0 \end{pmatrix} \begin{pmatrix} H_x \\ H_y \\ H_z \end{pmatrix} \quad \textbf{(A1a)}$$

$$\begin{bmatrix} B_x \\ B_y \\ B_z \end{bmatrix} = \begin{pmatrix} 0 & 0 & 0 \\ \zeta_i/c & 0 & 0 \\ 0 & 0 & 0 \end{pmatrix} \begin{pmatrix} E_x \\ E_y \\ E_z \end{pmatrix} + \mu_0 \begin{pmatrix} \mu_{r,i} & 0 & 0 \\ 0 & \mu_{r,i} & 0 \\ 0 & 0 & \mu_{r,i} \end{pmatrix} \begin{pmatrix} H_x \\ H_y \\ H_z \end{pmatrix} \quad \textbf{(A1b)}$$

, where $c$ is the light speed in vacuum; two studied materials are denoted by subscripts $i=1,2$, respectively; $\varepsilon$, $\xi$, $\zeta$, $\mu$ correspond to electromagnetic constitutive parameters; $E_x$, $D_x$, $H_y$, $B_y$ refer to the $x$-polarized electric field, $x$-polarized displacement field, $y$-polarized magnetic field, and $y$-polarized magnetic flux density, respectively. In this work, we assume $\xi_i = \zeta_i = i\kappa_i$, where $\kappa$ is the chirality parameter. If the studied wave is propagating along $z$-axis with $x$-polarized electric field and $y$-polarized magnetic field, the Eq. (A1a&A1b) is simplified into Eq. (1a&1b) in the main text. By substituting Eq. (A1) into Maxwell equation, one can readily obtain the eigen-equation of the bi-anisotropic material as

$$n_i \begin{bmatrix} D_x \\ B_y \end{bmatrix} = \mathcal{M}_i \sigma_x \begin{bmatrix} D_x \\ B_y \end{bmatrix} \quad \textbf{(A2)}$$

, where $n_i = k_i/\omega_i$ is the effective refractive index (eigen-value of the eigen-equation); $\sigma_x$ is the first Pauli matrix; and $\mathcal{M}_i$ is a material matrix expressed as

$$\mathcal{M}_i = \begin{bmatrix} c\varepsilon_0 \varepsilon_{r,i} & i\kappa_i \\ i\kappa_i & c\mu_0 \end{bmatrix}. \quad \textbf{(A3)}$$

By solving the Eq. (A2), the effective refractive index should be

$$n_i^\pm = i\kappa_i \pm \sqrt{\varepsilon_{r,i}} \qquad (A4)$$

, corresponding to the eigen-states $\bar{\Psi}_i^\pm = \begin{bmatrix} Z_i \\ \pm 1 \end{bmatrix}$, where the impedance is expressed as $Z_i = \sqrt{\dfrac{\varepsilon_0 \varepsilon_{r,i}}{\mu_0}}$.

Next, we study the material's symmetry by analyzing the material matrix. Take an example, if the material matrix satisfies $\mathcal{M} = \mathcal{M}^T$, implying the material is anomalous parity-time (*HPT*) symmetry, which represents the multiplication of the three operators— *H* (the Hermiticity), *P* (the inversion symmetry) and *T* (the time-reversal) [8]. To obtain a generalized understanding of the material's symmetry, we rewrite Eq. (A2) as $n\bar{\Psi} = \mathcal{M}\sigma_x \bar{\Psi}$. After implementing a general symmetry operator $E$, we have $n(E\bar{\Psi}) = \sigma_x (E\mathcal{M}E^{-1})(E\bar{\Psi})$. If the system is *E*-invariant, then $\mathcal{M} = E\mathcal{M}E^{-1}$. When $E = HPT$, the symmetry operator to wavefunction is considered as $E\bar{\Psi} = HPT\bar{\Psi} = \begin{bmatrix} -D_x & -B_y \end{bmatrix}$, essentially turning a right vector into a left one. We can obtain a new eigen-equation, $n\begin{bmatrix} -D_x & -B_y \end{bmatrix} = \sigma_x \begin{bmatrix} -D_x & -B_y \end{bmatrix} \mathcal{M}$, and perform a transpose operator to both side of the new equation, i.e., $n\begin{bmatrix} D_x \\ B_y \end{bmatrix} = \sigma_x \mathcal{M}^T \begin{bmatrix} D_x \\ B_y \end{bmatrix}$. Compared with the Eq. (A2), the material matrix satisfies $\mathcal{M} = \mathcal{M}^T$ so that the constitutive parameters fulfill $\xi = \zeta$. More specific results on other symmetries could be seen in Tab.1 in Ref. [8].

## Appendix B: Eigen-spectra under OBC and PBC

$\varepsilon_{r,i}$ and $\kappa_i$ respectively correspond to the relative permittivity and chirality parameter in the constitutive material marked by $i$. $i=1$ represents the isotropic materials with $\varepsilon_{r,1}=1$ and $\kappa_1=0$. $i=2(3)$ represents the bi-anisotropic material (i.e. non-Hermitian material) with relative permittivity $\varepsilon_{r,2}=4$ ($\varepsilon_{r,3}=4$) and chiral parameters $\kappa_2=-\kappa$ ($\kappa_3=\kappa$). Here, the specific values $\kappa=0$ and $\kappa=0.89$ correspond to Fig. 2(b,c,d) and Fig. 2(f,g,h), respectively. And $\kappa=0$, $\kappa=0.18$ and $\kappa=-0.18$ correspond to Fig. 3(b), (c) and (d), respectively. Note that when $\kappa=0$, the bi-anisotropic (non-Hermitian) material becomes the isotropic (Hermitian) one. The time durations for isotropic and bi-anisotropic time slabs are $t_1=T/3$ and $t_{2(3)}=2T/3$, respectively, where $T=1$ ns.

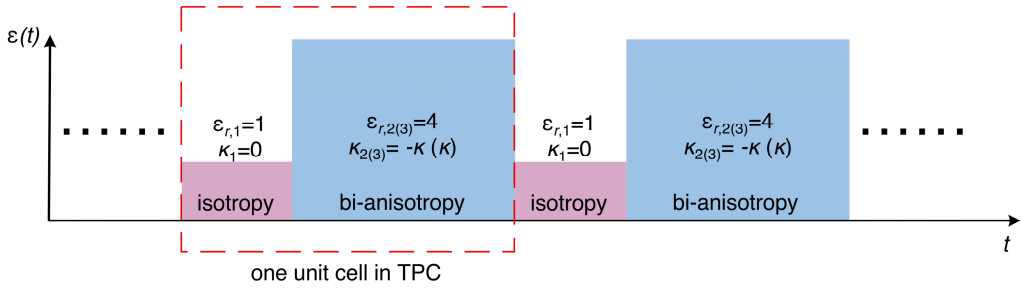

**Fig. A1: The time evolution of relative permittivity and chiral parameter in TPC.**

In this appendix, we show the derivation procedure of the eigen-spectra of temporal bulk and edge states. To illustrate this, we consider the same time varying structure as that in the main text (Fig. A2). Differently, transmission coefficients $T$ and reflection coefficients $R$ are introduced at each time interface between the isotropic materials and bi-anisotropic ones.

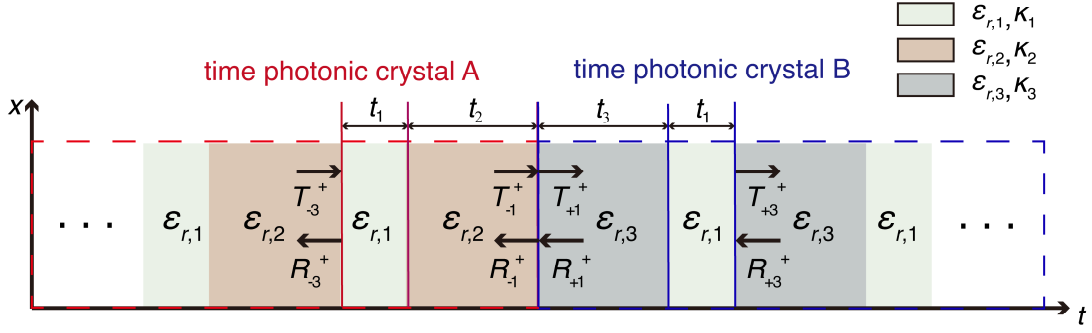

**Fig. A2: The schematic of time varying structure made of two TPCs.**

### (1) BZ and eigen-spectra of temporal bulk states under PBC

Without loss of generality, $t$ is given as the temporal duration inside a material. Therefore, we have the coefficients of forward propagating waves $T$ and coefficients of backward propagating waves $R$ on basis of the transfer matrix $S$:

$$\begin{bmatrix} T \\ R \end{bmatrix} = S \begin{bmatrix} T_1^+ \\ R_1^+ \end{bmatrix} = \begin{bmatrix} e^{-i\omega^+ t} & 0 \\ 0 & e^{-i\omega^- t} \end{bmatrix} \begin{bmatrix} T_1^+ \\ R_1^+ \end{bmatrix} \quad (A5)$$

Following this approach, we can calculate the electromagnetic field inside an isotropic and bi-anisotropic material, corresponding to the subscripts $i=1$ and $i=2(3)$, respectively.

$$\begin{bmatrix} D_x \\ B_y \end{bmatrix} = \begin{bmatrix} Z_i & Z_i \\ 1 & -1 \end{bmatrix} \begin{bmatrix} T \\ R \end{bmatrix} \quad (A6)$$

The transfer matrix across isotropic or bi-anisotropic materials can be expressed by

$$S_i = \begin{bmatrix} e^{-i\omega_i^+ t_i} & 0 \\ 0 & e^{-i\omega_i^- t_i} \end{bmatrix} \quad (A7)$$

, respectively; the superscript of plus and minus corresponds to the forward/backward propagating waves; $t_1$ and $t_{2(3)}$ are the temporal duration of the isotropic and bi-anisotropic materials, respectively; $\omega_1$ and $\omega_{2(3)}$ refers to the angular frequency of the isotropic and bi-anisotropic materials, respectively. We consider that the displacement field and magnetic flux density is continuous at the time interface between isotropic and bi-anisotropic materials, such that the transfer matrix from the bi-anisotropic material to isotropic material is indicated by

$$S_{1,2(3)} = \begin{bmatrix} Z_1 & Z_1 \\ 1 & -1 \end{bmatrix}^{-1} \begin{bmatrix} Z_{2(3)} & Z_{2(3)} \\ 1 & -1 \end{bmatrix} \tag{A8a}$$

The transfer matrix from the isotropic material to bi-anisotropic material is given by

$$S_{2(3),1} = \begin{bmatrix} Z_{2(3)} & Z_{2(3)} \\ 1 & -1 \end{bmatrix}^{-1} \begin{bmatrix} Z_1 & Z_1 \\ 1 & -1 \end{bmatrix} \tag{A8b}$$

$Z_{2(3)}$ and $Z_1$ are the impedances of bi-anisotropic materials and isotropic materials. Take one unit cell of TPC B as a typical example, we have

$$\begin{bmatrix} T_3^+ \\ R_3^+ \end{bmatrix} = \beta_B \begin{bmatrix} T_1^+ \\ R_1^+ \end{bmatrix} = S_{3,1} S_1 S_{1,3} S_3 \begin{bmatrix} T_1^+ \\ R_1^+ \end{bmatrix} = e^{-i\omega^\Sigma T} \begin{bmatrix} A_R & B_R \\ C_R & D_R \end{bmatrix} \begin{bmatrix} T_1^+ \\ R_1^+ \end{bmatrix} \tag{A9}$$

, where $\beta_B = e^{-i\omega_B T}$ is the generalized Bloch phase shift; $\omega_B$ is the Bloch angular frequency. By solving the Eq. (A9), we obtain the dispersion relationship of TPC B

$$\cos\left[(\omega_B - \omega^\Sigma)T\right] = \frac{\eta+1}{2}\cos\left(\omega_3^\Delta t_3 + \omega_1^\Delta t_1\right) + \frac{-\eta+1}{2}\cos\left(\omega_3^\Delta t_3 - \omega_1^\Delta t_1\right) \tag{A10}$$

and the scattering coefficients of eigen-function

$$\begin{bmatrix} T_1^+ \\ R_1^+ \end{bmatrix} = \begin{bmatrix} B_R \\ e^{-i(\omega_B - \omega^\Sigma)T} - A_R \end{bmatrix} \tag{A11}$$

, where $\omega_1^\Delta = \frac{\omega_1^+ - \omega_1^-}{2}$, $\omega_3^\Delta = \frac{\omega_3^+ - \omega_3^-}{2}$, and $\eta = \left(\frac{Z_3}{Z_1} + \frac{Z_1}{Z_3}\right)/2$. $T$ is the time period of TPC. $\omega^\Sigma T = \omega_1 t_1 + \omega_3 t_3$ where $\omega_1 = \frac{\omega_1^+ + \omega_1^-}{2}$, $\omega_3 = \frac{\omega_3^+ + \omega_3^-}{2}$. It is noted that the dispersion in TPC A has the same expression as the Eq. (A10), thus a general equation is given by Eq. (4). By solving Eq. (A10), we have the eigen-spectra for the TPC under PBC.

**(2) GBZ and eigen-spectra of temporal bulk states under OBC**

The eigen-spectra under PBC and OBC are different in TPC composed of non-Hermitian materials. To obtain the eigen-spectra under OBC, the open boundary condition should be taken into account. Based on Eq. (A6), the electric displacement inside the TPC takes a general form of

$$D_x = Z_i T_1^+ \beta_N^+ + Z_i R_1^+ \beta_N^-. \tag{A12}$$

, where the superscript of plus and minus corresponds to the forward/backward propagating direction; $\beta_N^\pm = e^{-i\omega_B^\pm NT}$ denote the generalized Bloch phase shift after $N$ time periods of TPC. Now we enforce the boundary condition of $D_x = 0$ at the temporal interface of TPC ($t=0$ and $t=NT$), then we should obtain

$$\beta_N^+ = \beta_N^- \text{ or } \omega_B^+ - \omega_B^- = \frac{2\pi m}{NT} \tag{A13}$$

, where $m$ is an integer. The condition can be simplified as

$$|\beta^+| = |\beta^-| \text{ or } \text{Im}(\omega_B^+) = \text{Im}(\omega_B^-) \tag{A14}$$

meaning the exponential decay of the forward and backward waves is the same. And we can find the OBC is fulfilled at an arbitrary temporal duration of TPC. By combining Eq. (A10) and (A14), we have the eigen-spectra and GBZ of the TPC under OBC.

**(3) Eigen-spectra of temporal edge states**

To solve the eigen-spectra of temporal edge states, we apply the temporal boundary conditions, i.e., the continuity of electrical displacement field and the magnetic flux density at the boundary, to above solved temporal bulk states at the interface between TPC A and TPC B as shown in Fig. A2:

$$\begin{pmatrix} Z_i & Z_i \\ 1 & -1 \end{pmatrix} \begin{bmatrix} T_{-1}^+ \\ R_{-1}^+ \end{bmatrix} = \begin{pmatrix} Z_i & Z_i \\ 1 & -1 \end{pmatrix} \begin{bmatrix} T_1^+ \\ R_1^+ \end{bmatrix} \tag{A15}$$

By solving the Eq. (A15), we have

$$\frac{T_1^+ + R_1^+}{T_{-1}^+ + R_{-1}^+} = \frac{T_1^+ - R_1^+}{T_{-1}^+ - R_{-1}^+} \tag{A16}$$

From Eq. (A14 and A11), one can solve for the eigen-values (wavevectors) of temporal edge states. These wavevectors form the eigen-spectra of temporal edge states.

## Appendix C: The penetration depth of temporal bulk and edge states

Based on the matrix transfer method, the temporal localization of bulk states could be realized in a straightforward way. To express the amplification or attenuation induced by chirality term, we multiply both sides of Eq. (A9) by a nonzero complex coefficient $e^{gT}$

$$\left(e^{gT}\beta_B\right) \begin{bmatrix} T_1^+ \\ R_1^+ \end{bmatrix} = S_{3,1} S_1 S_{1,3} \left(e^{gT} S_3\right) \begin{bmatrix} T_1^+ \\ R_1^+ \end{bmatrix} \tag{A17}$$

, where we suppose $\beta_B' = e^{gT}\beta_B = e^{-i(\omega_B + ig)T}$ and $S_3' = e^{gT} S_3$ to have a new material whose transfer matrix is $\begin{bmatrix} e^{-i(\omega_3^+ + \frac{igT}{t_3})t_3} & 0 \\ 0 & e^{-i(\omega_3^- + \frac{igT}{t_3})t_3} \end{bmatrix}$. Correspondingly, we have new angular frequencies $\omega_3'^{\pm} = \omega_3^{\pm} + \frac{igT}{t_3}$, new effective refractive indices $\frac{1}{n_3'^{\pm}} = (\frac{1}{\pm\sqrt{\varepsilon_{r,3}}} + \frac{igT}{t_3 kc})$, and a new Bloch angular frequency $\omega_B' = \omega_B + ig$. To obtain the amplification or attenuation rate $g$, we consider the new effective refractive indices are equal to the effective refractive indices in Eq. (A4). Without loss of generality, the

Eq. (A18) is also applicable in TPC A, thus we have

$$\frac{1}{n_{2(3)}^{'\pm}} = \frac{1}{i\kappa_{2(3)} \pm \sqrt{\varepsilon_{r,2(3)}}} = (\frac{1}{\pm\sqrt{\varepsilon_{r,2(3)}}} + \frac{igT}{t_{2(3)}kc}) \quad \textbf{(A18)}$$

By solving the Eq. (A18), we have $g = \frac{-kc\kappa_{2(3)}t_{2(3)}}{T\left(\varepsilon_{r,2(3)} + \kappa_{2(3)}^2\right)} \pm \frac{ikc\kappa_{2(3)}^2 t_{2(3)}}{T\sqrt{\varepsilon_{r,2(3)}}\left(\varepsilon_{r,2(3)} + \kappa_{2(3)}^2\right)}$.

Especially, we note that as $\text{Re}(g) \neq 0$, $|\beta_B'| \neq 1$, all bulk states become evanescent, resulting in the temporal localization of bulk states. The amplification or attenuation rate of bulk states is $\text{Im}(\omega_B') = \text{Re}(g)$. Correspondingly, the penetration depth of bulk states is

$$\delta_{bulk} = \frac{1}{|\text{Re}(g)|} = \left|\frac{T\left(\varepsilon_{r,2(3)} + \kappa_{2(3)}^2\right)}{kc\kappa_{2(3)}t_{2(3)}}\right| \quad \textbf{(A19)}$$

Next, we consider the penetration depth of edge states. In TPC composed of purely Hermitian materials, the initial penetration depth of edge states can be derived from the Eq. (A10). Substituting $k = \frac{k_0}{2} = 3\pi/(2cT)$ into the Eq. (A10), we have the Bloch angular frequency of edge states $\omega_0 = -\frac{\pi}{T} + ig_0$, where $g_0 = \text{acosh}(\eta/T)$ is a nonzero real number. The initial amplification or attenuation rate of edge states is $\text{Im}(\omega_0) = g_0$. The initial penetration depth of edge states is

$$\delta_0 = \frac{1}{|g_0|} \quad \textbf{(A20)}$$

And the temporal localization of bulk states, induced by bi-anisotropic materials, plays a key role in offsetting the initial penetration depth of edge states. In other words, the overall amplification or attenuation rate of edge states is $\text{Im}(\omega_0) \pm \text{Im}(\omega_B') = g_0 \pm \text{Re}(g)$ by taking into account the localization of bulk states. Correspondingly, the final penetration depth of edge states is

$$\delta_{edge} = \frac{1}{|g_0 \pm \mathrm{Re}(g)|} = \frac{1}{\left|\frac{1}{\delta_0} \pm \left|\frac{kc\kappa_{2(3)}t_{2(3)}}{T\left(\varepsilon_{r,2(3)} + \kappa_{2(3)}^2\right)}\right|\right|} \quad \text{(A21)}$$

# Appendix D: The experimental realization of time-variant bi-anisotropy in the material

In this appendix, we discuss the experimental realization of the bi-anisotropic material proposed in this paper. In Ref. [40], the bi-anisotropic media are generally classified into four categories, chiral, Tellegen, moving, and omega. Among these, the moving medium is the only one meeting the constitutive relation in Eq. (1) and material matrix in Eq. (3), given by

$$D_x = \varepsilon_0 \varepsilon_{r,i} E_x - (V/c) H_y \quad \text{(A22a)}$$

$$B_y = -(V/c) E_x + \mu_0 H_y \quad \text{(A22b)}$$

Compared with Eq. (1), when $V = -i\kappa_i$, such a medium is equivalent to the one we proposed in this paper. Ref. [41] presents the details to realize the complex values of velocity $V$, by periodically loading nonreciprocal gyrators to transmission lines, as illustrated in the Fig. A3(b). Such a composite structure can mimic the electromagnetic phenomena in moving media in a stationary reference frame. Note that the velocity value $V$ depends on the mutual coupling coefficient and resistance in the gyrator, which allows us to flexibly adjust the velocity value. By using voltage-controlled radio switches [i.e., component S in Fig. A3(b)], one can switch the bi-anisotropic material to isotropic material periodically in time. Moreover, the permittivity is also tunable in time by paralleling the capacitor array with voltage-controlled radio switches. Such a system allows us to modulate permittivity $\varepsilon_{r,i}$ and chiral parameter $V = -i\kappa_i$ periodically in time [Fig. A3(a)].

In addition to the transmission-line, such a bi-anisotropic material can also be designed based on the moving chiral meta-atoms [51] and an active meta-surface [52].

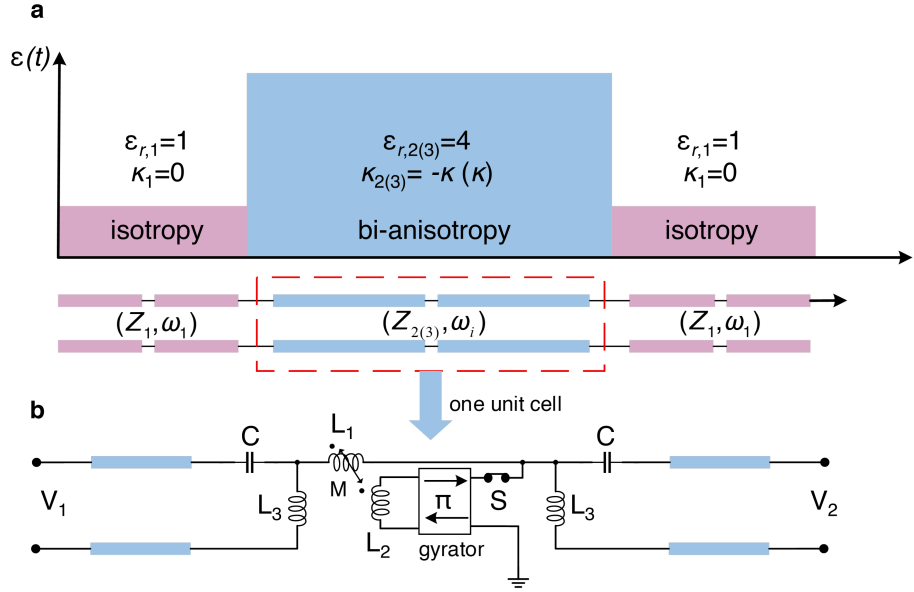

**Fig. A3: The experimental design of time photonic crystals composed of bi-anisotropic and isotropic materials. (a) Periodical transmission-line metamaterial with time modulation on permittivity and chiral parameter. (b) The schematic of one unit cell in transmission-line metamaterial.**

# Appendix E: The consideration of frequency-dependent chirality parameter

For practical considerations, we investigate the influence of material dispersions on the presented results, i.e., the broadband temporal localization and delocalized temporal edge states. In Fig. A3, the velocity parameter of bi-anisotropic material takes the form as $V = -i\kappa_i$. In realistic transmission-line system, the $\kappa_i$ should satisfy the Lorentz model as plotted in Fig. A4(a), where the resonant frequency is 1.02 GHz. As a result, the delocalized edge state emerges at $k=1.506k_0$ (determined by combining Eq. (8) and Lorentz dispersion of $\kappa_i$), as illustrated in Fig. A4(c). Moreover, when the resonant frequency of Lorentz model is tuned to be 0.75GHz, apparently, we could also observe the broadband temporal localization in Fig. A4(b) in broadband wavevectors from $k=0.612k_0$ to $k=1.396k_0$ $k=1.396k_0$.

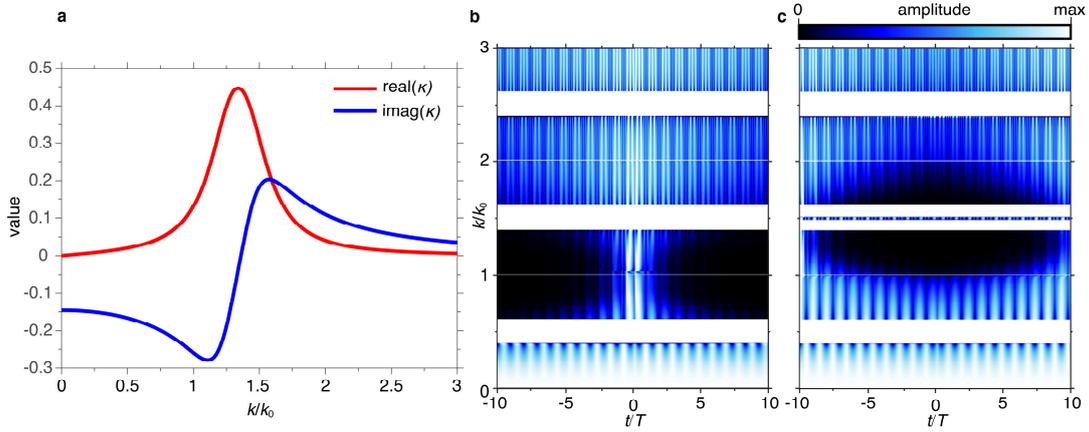

**Fig. A4: The broadband temporal localization and delocalized temporal edge states with frequency-dependent parameter. (a) The dispersion of chirality parameter. In (a), the red/blue solid line represents the real/imaginary part of chirality parameter. (b&c) Field distributions of electric displacement in the time varying structure with frequency-dependent chirality parameter. In (b), the resonant frequency is located at 0.75GHz. In (c), the resonant frequency is located at 1.02GHz.**

# Appendix F: The influence of time disorders on broadband temporal localization and delocalized temporal edge states

The gyrator-based transmission-line metamaterials provide a realistic platform to mimic the bi-anisotropic material distributed periodically in time. As the operating frequency increases, the time scale of this configuration would be reduced to nanosecond-scale and even a lower level. In this appendix, we investigate the influence of disorders on time scale. To start with, we introduce disorders in each temporal slab by randomly varying the duration of each time slab within the range of $[(1-\rho)t_i, (1+\rho)t_i]$, where the disorder rate $\rho$ is defined as the maximum variation strength of time in each slab of TPCs over their initial values. Compared with the results in Fig. 3, though the Fig. A5(a) demonstrates reduction of temporal localization strength of bulk states, we could still observe the broadband temporal localization with $\rho = 0.3$. However, the delocalized edge states are more sensitive to disorders and dissipate when $\rho = 0.1$, as illustrated in Fig. A5(b).

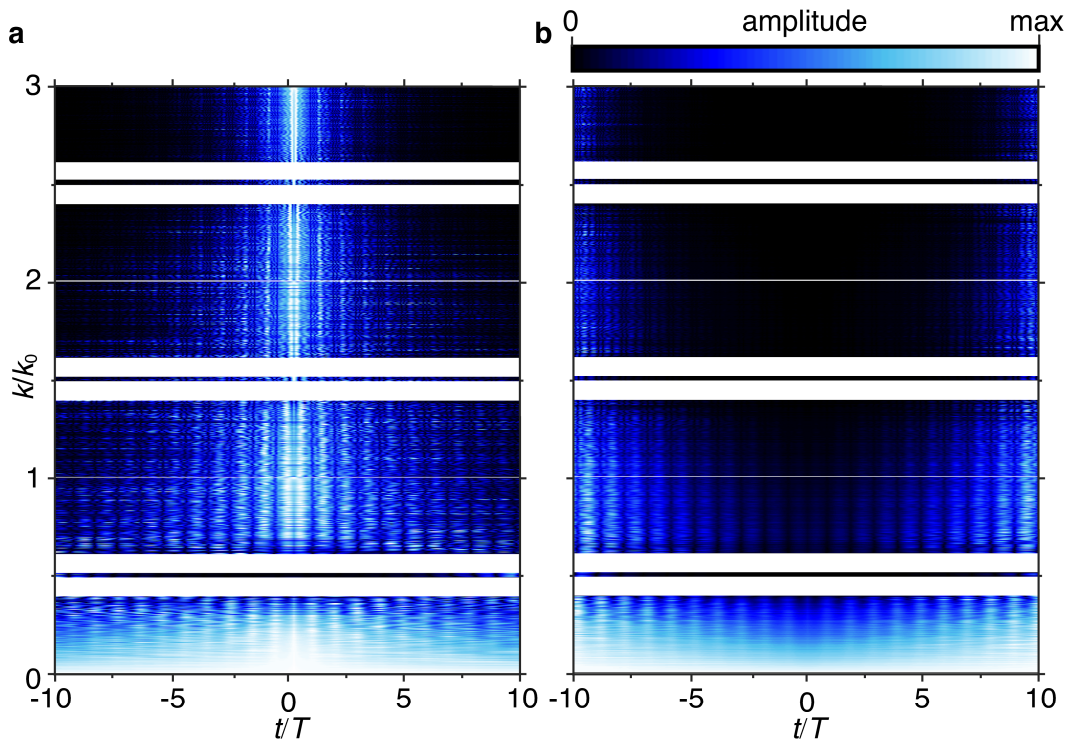

**Fig. A5. The impact of disorders on the temporal bulk and edge states. (a) Field distributions of electric displacement in the time varying structure with $\kappa_2 = -0.18$, $\kappa_3 = 0.18$, $\rho = 0.3$. (b) Field distributions of electric displacement in the time varying structure with $\kappa_2 = 0.18$, $\kappa_3 = -0.18$, $\rho = 0.1$.**

# Appendix G: The effect of material absorption on broadband temporal localization and delocalized temporal edge states

In this appendix, we investigate the effect of material absorption on main results, i.e., the broadband temporal localization and delocalized temporal edge states. The absorption is considered by introducing the imaginary part of permittivity. Namely, instead of fixing $\varepsilon_{r,2(3)} = 4$, a nonzero positive imaginary part ($\varepsilon_{r,2(3)} = 4 + \varepsilon_i * i$) is introduced onto the TPC A and TPC B in Fig. 3(a).

Firstly, we find that broadband temporal localization of bulk states is relatively robust against the material absorption. To be specific, the broadband temporal localization is still preserved if $\varepsilon_i = 0.2$, while it disappears if $\varepsilon_i = 0.6$ [see Fig. A6(b&c)].

Secondly, we notice that delocalized edge sates are more sensitive to material absorption. That is, it suddenly disappears if a tiny loss $\varepsilon_i = 0.01$ is considered in Fig. A7(b).

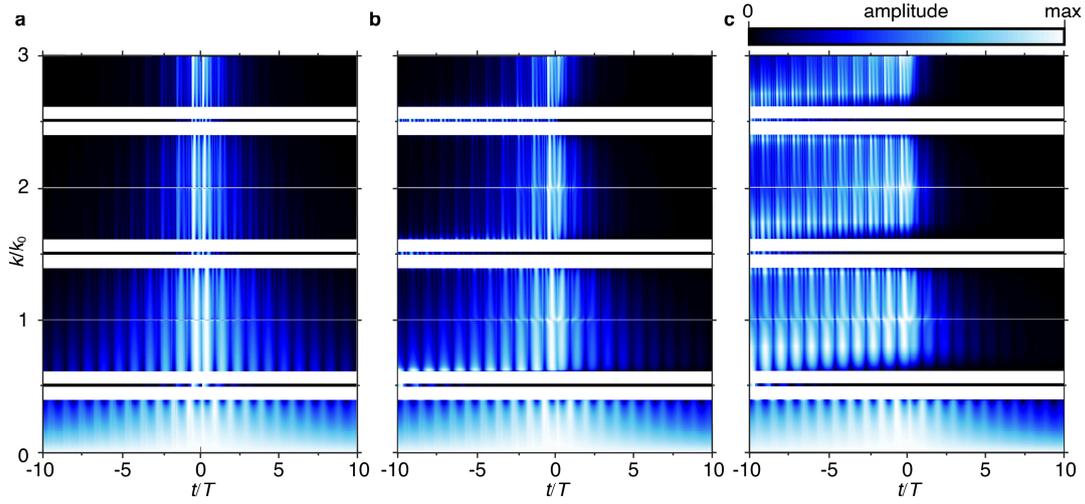

Fig. A6. The impact of absorption on the temporal bulk states. (a,b&c) Field distributions of electric displacement in the time varying structure with same chirality parameters $\kappa_2 = -0.18$, $\kappa_3 = 0.18$. The permittivity is different in (a,b&c) where (a) $\varepsilon_{r,2(3)} = 4$ (b) $\varepsilon_{r,2(3)} = 4 + 0.2*i$ (c) $\varepsilon_{r,2(3)} = 4 + 0.6*i$.

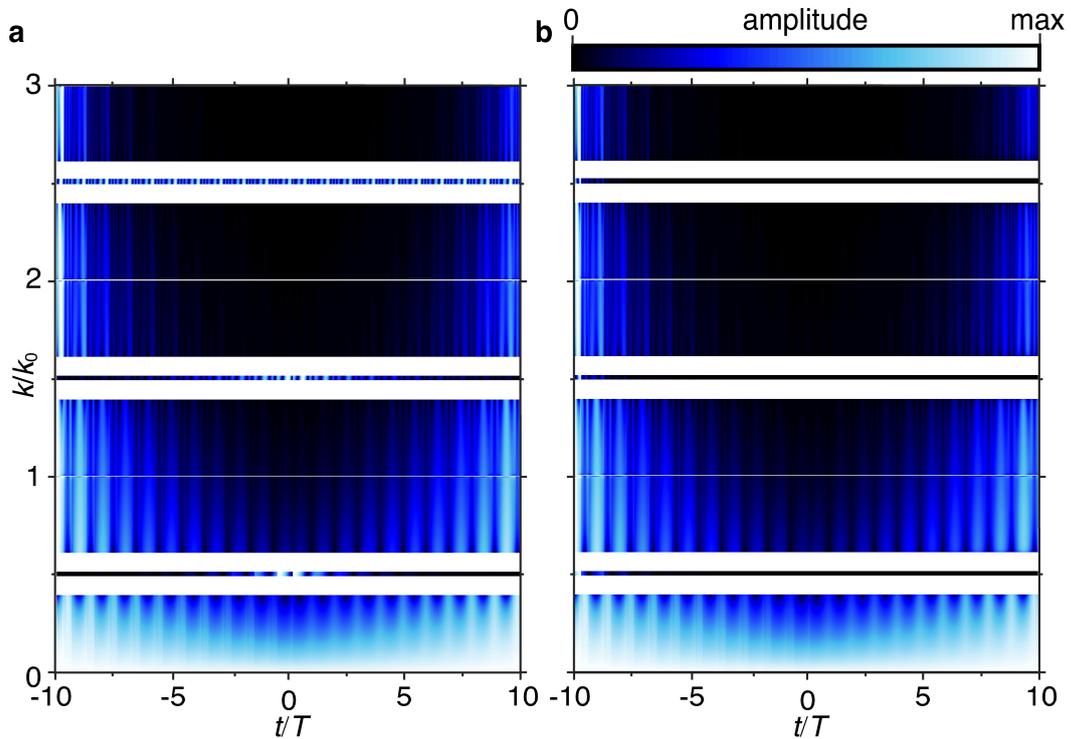

Fig. A7. The impact of absorption on the temporal edge states. (a&b) Field distributions of electric displacement in the time varying structure with same

chirality parameters $\kappa_2 = 0.18$, $\kappa_3 = -0.18$. The permittivity in (a) is $\varepsilon_{r,2(3)} = 4$ and that one in (b) is $\varepsilon_{r,2(3)} = 4 + 0.01*i$.

## Appendix H: The space-time dynamical description of temporal modes

To further reveal the consequence of temporal modes, we also analytically investigate the wave behaviors in the interaction between time photonic crystals and an incident pulse with a finite temporal width. We assume that an incident pulse takes the form $E_y = \text{rect}(\frac{t}{\tau})e^{-i\omega_0 t}$ in the time dimension, where $\text{rect}(x)$ is the rectangular function, $\tau$ is the temporal width of the incident pluses, and $\omega_0$ is the central frequency of incidence. By exploiting the plane-wave expansion method, we have $E_y = \tau \int \text{sinc}(\tau \frac{\omega - \omega_0}{2\pi}) e^{-i\omega t} d\omega$, where $\text{sinc}(x) = \sin(\pi x)/(\pi x)$. Now, one can apply the temporal boundary conditions for such an incidence to determine the electromagnetic response of the temporal modes in TPC. In Fig. A8&A9, the TPC1 and TPC2 are built as the one in Fig.3(a). An incident beam will be split into two wave beams, the transmitted and reflected waves in the TPC. When $\omega_0 = \pi/(4*T)$, where $T = 1(\text{ns})$, the transmitted and reflected modes will localize in the TPC, as shown in Fig. A8 (a)&(b). When $\omega_0 = 3\pi/(2*T)$ in Fig. A9(a)&(b), the transmitted and reflected modes (except edge modes) will demonstrate the amplification along the time domain.

Our analytical calculation reflects that in the TPC1, an incident beam will couple to two wave beams, one propagating in the positive *z* direction, and the other in the negative *z* direction. In the TPC2, the two wave beams keep propagating without scattering. Through TB2, the two wave beams split again, each coupling back to a forward and backward beam. As a result, in the Fig. A8, four different wave beams emerge in air. However, once the central frequency of incidence falls within the bandgap, only two wave beams emerge for an exponential amplification of energy as time progresses, as shown in Fig. A9.

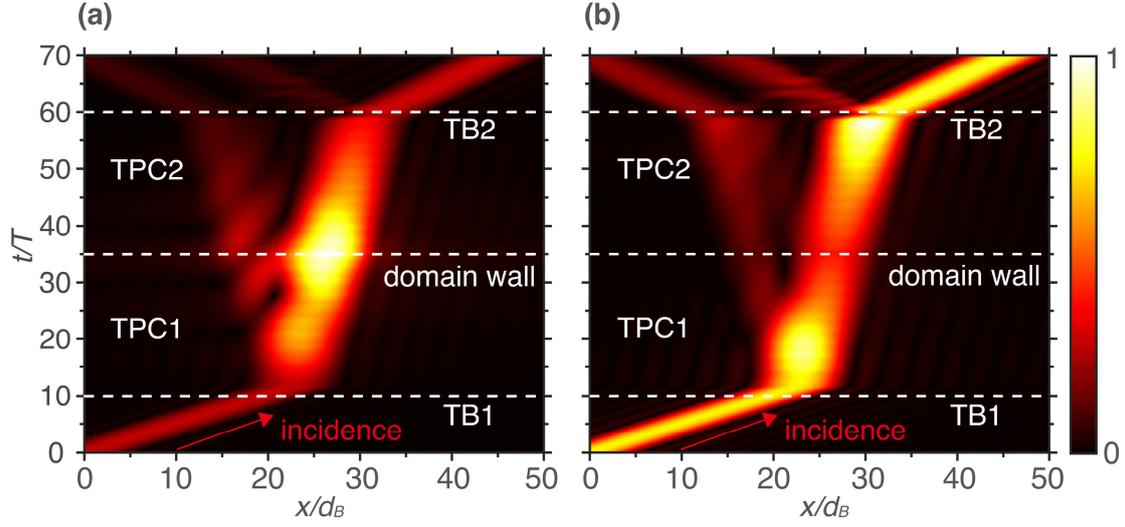

**Fig. A8: Space–time description of $|D(x,t)|$ on band propagation. The TPC1 and TPC2 are constructed as the one in Fig.3(a). The temporal length of TPC1 is $25*T$, and that of TPC2 is also $25*T$, where $T=1(\text{ns})$. The central frequency of incidence is $\omega_0 = \pi/(4*T)$. The white dashed line marks the start time (TB1) and the end time (TB2) of the TPC. (a) The chirality parameters in TPC are $\kappa_1=0$, $\kappa_2=-0.18$ and $\kappa_3=0.18$, corresponding to the parametric setting in the Fig.3(c); the transmitted and reflected modes are localized at the domain wall, corresponding to the localized bulk sates in the Fig.3(c). (b) The chirality parameters in TPC are $\kappa_1=0$, $\kappa_2=0.18$ and $\kappa_3=-0.18$, corresponding to the parametric setting in the Fig.3(d). the transmitted and reflected modes are localized at the TB1 and TB2, corresponding to the localized bulk sates in the**

Fig.3(d).

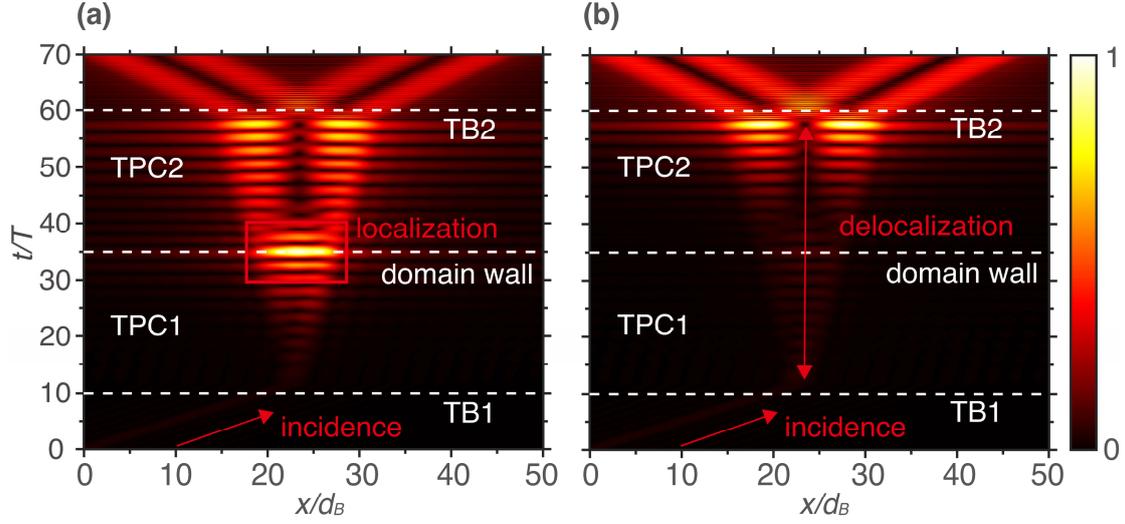

**Fig. A9:** Space–time description of $|D(x,t)|$ on gap propagation. The TPC1 and TPC2 are constructed as the one in Fig.3(a). The temporal length of TPC1 is $25*T$, and that of TPC2 is also $25*T$, where $T=1(\text{ns})$. The central frequency of incidence is $\omega_0 = 3\pi/(2*T)$. The white dashed line marks the start time (TB1) and the end time (TB2) of the TPC. (a) The chirality parameters in TPC are $\kappa_1=0$, $\kappa_2=-0.18$ and $\kappa_3=0.18$, corresponding to the parametric setting in the Fig.3(c);the transmitted and reflected modes are amplified along time, whereas the edge modes localize at domain wall, corresponding to the localized edge sates in the Fig.3(c). (b) The chirality parameters in TPC are $\kappa_1=0$, $\kappa_2=0.6$ and $\kappa_3=-0.6$; the transmitted and reflected modes are amplified along time, whereas the edge modes delocalize, corresponding to the delocalized edge sates in the Fig.3(d).

# Appendix I: The phenomena unique to time photonic crystals

Some phenomena unique to temporal boundaries are listed as follows:

(1) Temporal bandgaps, as revealed by many previous works, enable both amplification and attenuation of waves, while spatial bandgaps only allow attenuation of waves. As shown in Fig. A8&A9, inside bandgap of TPC, the eigen modes are amplified with time increasing.

(2) The behaviors (phases, magnitudes, attenuation and amplification rates) of temporal modes are independent on the upcoming interface of time photonic crystals. However, the behaviors of spatial modes highly depend on both the initial and final interfaces of space photonic crystals. In other words, the revealed temporal modes, including both the localized and delocalized ones, in this work, are more robust against the structural deformation. As shown in Fig. A10(b), before the upcoming temporal interface $t = 50T$, the waveforms corresponding to TPC1 (red solid) and TPC2 (blue dashed) are essentially indistinguishable. In sharp contrast, before the spatial interface $x = 50D$, Fig. A10(a) shows that the spatial fields in SPC1 (red solid) and SPC2 (blue dashed) differ with each other markedly.

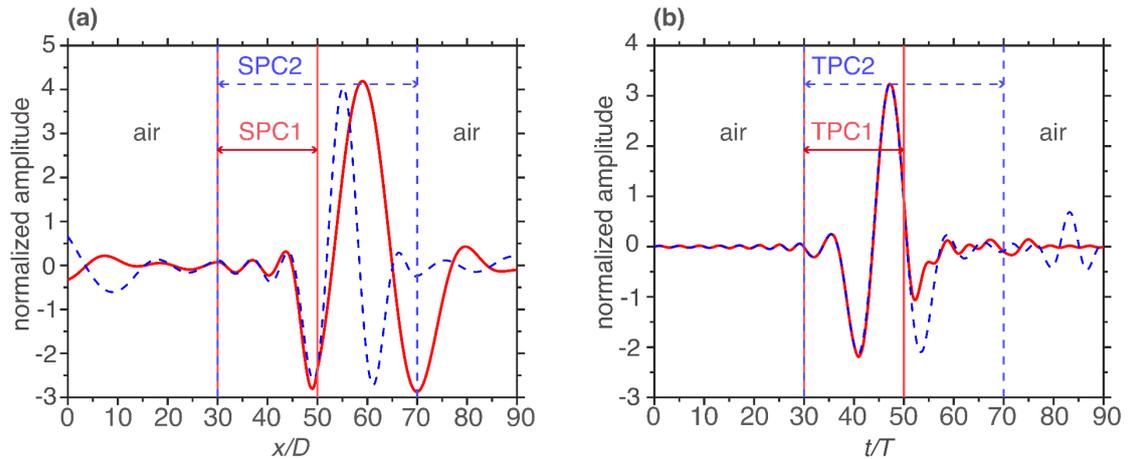

**Fig. A10: Comparison between spatial and temporal photonic crystals with different terminal interfaces. (a) Normalized field amplitude along position $x/D$**

for two space photonic crystals, SPC1 (red solid) and SPC2 (blue dashed), embedded in air; the SPC1 and SPC2 share the same entrance interface ($x = 30D$) but correspond to different terminal interface, $x_1 = 50D$ and $x_2 = 70D$, where $D = 0.1 \,(\mathrm{m})$. (b) Normalized field amplitude versus normalized time $t/T$ for two time photonic crystals, TPC1 (red solid) and TPC2 (blue dashed), with the same initial temporal interface ($t = 30T$); but TPC1 and TPC2 are corresponding to different upcoming temporal interfaces, $t_1 = 50T$ and $t_2 = 70T$, where $T = 1 \,(\mathrm{ns})$; the background material is air.

**References**

[1] C. E. Rüter, K. G. Makris, R. El-Ganainy, D. N. Christodoulides, M. Segev, and D. Kip, Observation of parity–time symmetry in optics, Nat. Phys. **6**, 192–195 (2010).

[2] V. M. Martinez Alvarez, J. E. Barrios Vargas, and L. E. F Foa Torres, Non-Hermitian robust edge states in one dimension: anomalous localization and eigenspace condensation at exceptional points, Phys. Rev. B **97**, 121401 (2018).

[3] Y. Ashida, Z. Gong, and M. Ueda, Non-Hermitian physics, Adv. Phys. **69**, 249-435 (2020).

[4] E. J. Bergholtz, J. C. Budich, and F. K. Kunst, Exceptional topology of non-Hermitian systems, Rev. Mod. Phys. **93**, 015005 (2021).

[5] F. Song, S. Yao, and Z. Wang, Non-Hermitian skin effect and chiral damping in open quantum systems, Phys. Rev. Lett. **123**, 170401 (2019).

[6] P. C. Cao, Y. Li, Y. Peng, M. Qi, W. Huang, P. Li, and X. Zhu, Diffusive skin effect and topological heat funneling, Commun. Phys. **4**, 230 (2021).

[7] S. Weidemann, M. Kremer, T. Helbig, T. Hofmann, A. Stegmaier, M. Greiter, R. Thomale, and A. Szameit, Topological funneling of light, Science **368**, 6488 (2020).

[8] Q. Yan, H. Chen, and Y. Yang, Non-Hermitian skin effect and delocalized edge states in photonic crystals with anomalous parity-time symmetry, arXiv:2111.08213 (2021).

[9] M. Gao, C. Sheng, Y. Zhao, R. He, L. Lu, W. Chen, K. Ding, S. Zhu, and H. Liu, Quantum walks of correlated photons in non-Hermitian photonic lattices, Phys. Rev. B **110**, 094308 (2024).

[10] K. Kawabata, T. Numasawa, and S. Ryu, Entanglement phase transition induced by the non-Hermitian skin effect, Phys. Rev. X **13**, 021007 (2023).

[11] S. Yao and Z. Wang, Edge states and topological invariants of non-Hermitian systems, Phys. Rev. Lett. **121**, 086803 (2018).

[12] F. Song, S. Yao, and Z. Wang, Non-Hermitian topological invariants in real space, Phys. Rev. Lett. **123**, 246801 (2019).

[13] K. Yokomizo and S. Murakami, Non-Bloch band theory of non-Hermitian systems, Phys. Rev. Lett. **123**, 066404 (2019).

[14] S. Longhi, Probing non-Hermitian skin effect and non-Bloch phase transitions, Phys. Rev. Res. **1**, 023013 (2019).

[15] L. Xiao, T. Deng, K. Wang, G. Zhu, Z. Wang, W. Yi, and P. Xue, Non-Hermitian bulk–boundary correspondence in quantum dynamics, Nat. Phys. **16**, 761–766 (2020).

[16] T. Helbig, T. Hofmann, S. Imhof, M. Abdelghany, T. Kiessling, L. W. Molenkamp, C. H. Lee, A. Szameit, M. Greiter, and R. Thomale, Generalized bulk–boundary correspondence in non-Hermitian topolectrical circuits, Nat. Phys. **16**, 747–750 (2020).

[17] A. McDonald and A. A. Clerk, Exponentially-enhanced quantum sensing with non-Hermitian lattice dynamics, Nat. Commun. **11**, 5382 (2020).

[18] L. Xiao, T. Deng, K. Wang, Z. Wang, W. Yi, and P. Xue, Observation of non-Bloch parity-time symmetry and exceptional points, Phys. Rev. Lett. **126**, 230402 (2021).

[19] W. Xue, M. Li, Y. Hu, F. Song, and Z. Wang, Simple formulas of directional amplification from non-Bloch band theory, Phys. Rev. B **103**, L231408 (2021).

[20] H. Zirnstein, G. Rafael, and B. Rosenow, Bulk-boundary correspondence for non-Hermitian Hamiltonians via green functions, Phys. Rev. Lett. **126**, 216407 (2021).

[21] Y. Yang, H. Hu, L. Liu, Y. Yang, Y. Yu, Y. Long, X. Zheng, Y. Luo, Z. Li, and F. J. Garcia-Vidal, Topologically protected edge states in time photonic crystals with chiral symmetry, ACS Photon. **12**, 2389-2396 (2025).

[22] Y. Yu, et al., Generalized coherent wave control at dynamic interfaces, Laser Photonics Rev. **19,** 202400399 (2024).

[23] S. Yin, Y. Wang, and A. Alù, Temporal optical activity and chiral time-interfaces, Opt. Express **30**, 47933-47941 (2022).

[24] X, Wang, G. Ptitcyn, V. S. Asadchy, A. Díaz-Rubio, M. S. Mirmoosa, S. Fan, and S. A. Tretyakov, Nonreciprocity in bianisotropic systems with uniform time modulation, Phys. Rev. Lett. **125**, 266102 (2020).

[25] D. L. Sounas and A. Alù, Non-reciprocal photonics based on time modulation, Nat. Photon. **11**, 774–783 (2017).

[26] H. Li, S. Yin, and A. Alù, Nonreciprocity and faraday rotation at time interfaces, Phys. Rev. Lett. **128**, 173901 (2022).

[27] L. Stefanini, et al., Time-varying metasurfaces for efficient surface-wave coupling to radiation and frequency conversion, Laser Photonics Rev. **18**, 2400315 (2024).

[28] H. Moussa, G. Xu, S. Yin, E. Galiffi, Y. Ra'di, and A. Alù, Observation of temporal reflection and broadband frequency translation at photonic time interfaces, Nat. Phys. **19**, 863–868 (2023).

[29] N. Wang and G. P. Wang, Broadband frequency translation by space–time interface with weak permittivity temporal change, Opt. Lett. **48**, 4436 (2023).

[30] X. Wang, M. S. Mirmoosa, V. S. Asadchy, C. Rockstuhl, S. Fan, and S. A. Tretyakov, Metasurface-based realization of photonic time crystals, Sci. Adv. **9**, eadg7541 (2023).

[31] X. Wang, P. Garg, M. S. Mirmoosa, A. G. Lamprianidis, C. Rockstuhl, and V. S. Asadchy, Expanding momentum bandgaps in photonic time crystals through resonances, Nat. Photonics **19**, 149–155 (2025).

[32] H. Li, S. Yin, H. He, J. Xu, A. Alù, and B. Shapiro, Stational charge radiation in anisotropic photonic time crystals, Phys. Rev. Lett. **130**, 093803 (2023).

[33] M. Lyubarov, Y. Lumer, A. Dikopoltsev, E. Lustig, Y. Sharabi, and M. Segev, Amplified emission and lasing in photonic time crystals, Science **377**, 425–428 (2022).

[34] A. Dikopoltsev, et al., Light emission by free electrons in photonic time-crystals, Proc. Natl. Acad. Sci. USA **119**, e2119705119 (2022).

[35] X. Gao, X. Zhao, X. Ma, and T. Dong, Free electron emission in vacuum assisted by photonic time crystals, J. Phys. D: Appl. Phys. **57**, 315112 (2024).

[36] E. Galiffi, et al., Photonics of time-varying media, Adv. Photon. **4**, 014002 (2022).

[37] G. Liu, et al., Photonic axion insulator, Science **387**, 162-166 (2025).

[38] M. H. M. Mostafa, M. S. Mirmoosa, and S. A. Tretyakov, Spin-dependent phenomena at chiral temporal interfaces, Nanophotonics **12**, 2881-2889 (2023).

[39] D. Yang, J. Xu, and D. H. Werner, A generalized temporal transfer matrix method and its application to modeling electromagnetic waves in time-varying chiral media, Appl. Phys. Lett. **122,** 251102 (2023).

[40] M. S. Mirmoosa, M. H. Mostafa, A. Norrman, and S. A. Tretyakov, Time interfaces in bianisotropic media, Phys. Rev. Research **6**, 013334 (2024).

[41] J. Vehmas, S. Hrabar, and S. Tretyakov, Transmission lines emulating moving media, New J. Phys. **16**, 093065 (2014).

[42] J. R. Zurita-Sánchez, P. Halevi, and J. C. Cervantes-González, Reflection and transmission of a wave incident on a slab with a time-periodic dielectric function ε(t), Phys. Rev. A **79**, 053821 (2009).

[43] M. Li, J. Liu, X. Wang, W. Chen, G. Ma, and J. Dong, Topological temporal boundary states in a non-Hermitian spatial crystal, arXiv:2306.09627 (2023).

[44] M. M. Asgari, P. Garg, X. Wang, M. S. Mirmoosa, C. Rockstuhl, and V. Asadchy, Theory and applications of photonic time crystals: a tutorial, Adv. Opt. Photon. **16**, 958-1063 (2024).

[45] T. E. Lee, Anomalous edge state in a non-Hermitian lattice, Phys. Rev. Lett. **116**, 133903 (2016).

[46] Y. Yi and Z. Yang, Non-Hermitian skin modes induced by onsite dissipations and chiral tunneling effect, Phys. Rev. Lett. **125**, 186802 (2020).

[47] L. Li, C. H. Lee, and J. Gong, Topological Switch for Non-Hermitian skin effect in cold-atom systems with loss, Phys. Rev. Lett. **124**, 250402 (2020).

[48] Y. Li, C. Liang, C. Wang, C. Lu, and Y.-C. Liu, Gain-loss induced hybrid skin-topological effect, Phys. Rev. Lett. **128**, 223903 (2022).

[49] W.-T. Xue, Y.-M. Hu, F. Song, and Z. Wang, Non-Hermitian edge burst, Phys. Rev. Lett. **128**, 120401 (2022).

[50] R. Okugawa, R. Takahashi and K. Yokomizo, Non-Hermitian band topology with generalized inversion symmetry, Phys. Rev. B **103**, 205205 (2021).

[51] M. S. Mirmoosa, Y. Ra'di, V. S. Asadchy, C. R. Simovski and S. A. Tretyakov, Polarizabilities of nonreciprocal bianisotropic particles, Phys. Rev. Applied **1**, 034005 (2014).

[52] K. Achouri, B. A. Khan, C. Caloz, V. Asadchy and S. Tretyakov, Synthesis of a nongyrotropic nonreciprocal metasurface as an equivalent to a moving medium, (2016 IEEE International Symposium on Antennas and Propagation (APSURSI), Fajardo, 2016), pp. 371-372.

[53] W. Y. Cui, J. Zhang, Y. Luo, X. Gao and T. J. Cui, Dynamic switching from coherent perfect absorption to parametric amplification in a nonlinear spoof plasmonic waveguide, Nat. Commun. **15**, 2824 (2024).